\documentclass[12pt,preprint]{aastex}

\slugcomment{Submitted to The Astronomical Journal}

\begin{document}

%
% Definitions.
%
\newcommand {\hi} {\ion{H}{1}\,\,}
\newcommand {\hII} {\ion{H}{2}\,\,}
\newcommand {\ha} {H$\alpha$\,\,}
\newcommand {\kms} {\,km\,s$^{-1}$\,}
\newcommand {\M} {\mbox{${\cal M}$}}
\newcommand {\mhi} {\M$_{HI}$\,}
\newcommand {\mhil} {$M_{HI} / L_B$\ }
\newcommand {\msol} {\M$_\odot$\,}
\newcommand {\lsol} {\L$_\odot$\,}

\title{\hi Detection in two Dwarf S0 Galaxies in Nearby Groups:
ESO384-016 and NGC 59}

\author{Sylvie F. Beaulieu}
\affil{D\'epartement de Physique, de G\'enie Physique et d'Optique
and Observatoire du mont M\'egantic, 
Universit\'e Laval, Qu\'ebec, Qc, G1K 7P4, Canada}
\email{sbeaulieu@phy.ulaval.ca}

\author{Kenneth C. Freeman}
\affil{Research School of Astronomy and Astrophysics, Mount Stromlo
Observatory, Weston Creek, ACT 2611, Australia}
\email{kcf@mso.anu.edu.au}

\author{Claude Carignan}
\affil{D\'epartement de Physique and Observatoire du mont M\'egantic, 
Universit\'e de Montr\'eal, C.P. 6128, Succ. Centre-ville, Montr\'eal,
Qc, H3C 3J7, Canada}
\email{Claude.Carignan@umontreal.ca}

\author{Felix J. Lockman}
\affil{National Radio Astronomy Observatory, P.O. Box 2, Green Bank,
WV 24944, USA}
\email{jlockman@nrao.edu}

\and

\author{Helmut Jerjen}
\affil{Research School of Astronomy and Astrophysics, Mount Stromlo
Observatory, Weston Creek, ACT 2611, Australia}
\email{jerjen@mso.anu.edu.au}

\begin{abstract}
An \hi survey of 10 dE/dS0 galaxies in the
nearby Sculptor and Centaurus A groups was made using the Australia
Telescope Compact Array (ATCA). The observed galaxies have
accurate distances derived by Jerjen et al (1998; 2000b)
using the surface brightness fluctuation
technique. Their absolute magnitudes are in the range $-9.5 > M_B
> -15.3$. Only two of the ten galaxies were detected at our
detection limit ($\sim 1.0 \times 10^6$ \msol for the Centaurus group
and $\sim 5.3 \times 10^5$ \msol for the Sculptor group), the two dS0
galaxies ESO384-016 in the Centaurus A Group and NGC 59 in the
Sculptor Group, with \hi masses of $6.0 \pm 0.5 \times 10^6$ \msol and
$1.4 \pm 0.1 \times 10^7$ \msol respectively. Those two detections were 
confirmed using the Green Bank Telescope. These small \hi reservoirs
could fuel future generations of low level star formation
and could explain the bluer colors seen at the center of the
detected galaxies. Similarly to what is seen with the Virgo dEs, the
two objects with \hi appear to be on the outskirt of the groups.
\end{abstract}


\section{Introduction}

The nature and origin of the fainter dwarf elliptical galaxies, and
their relationship to the brighter ellipticals, remains poorly
understood. Kormendy (1985) examined the structural scaling laws
for the two classes of ellipticals (eg the correlations between
absolute magnitude, core radius and effective surface brightness)
and found a discontinuity between the scaling laws for the dwarf
and brighter ellipticals. This discovery contributed significantly
to the view that dwarf and giant ellipticals are formed by different
processes. The issue remains controversial, however. Some authors
(eg Jerjen \& Binggeli 1997 ; Graham \& Guzm\'an 2003) argue that the 
scaling law discontinuities
are not fundamental but come from systematic changes of the surface
brightness profiles with galaxy luminosity, such as the systematic
changes of the index $n$ of the empirical S\'ersic law changes with
luminosity.

Independent of the structural issues, there do appear to be three
dynamical classes of ellipticals. The internal kinematics of
ellipticals show some unambiguous changes with luminosity, with the
breaks between classes located at $M_B \approx -17.5$ and $M_B
\approx -19.5$ ($H_\circ = 75$ \kms Mpc$^{-1}$). At $M_B
\approx -17.5$, there is a marked transition in the dynamical
properties between the intermediate luminosity ellipticals and the
fainter dwarf ellipticals.  The intermediate ellipticals ($-17.5 > M_B >
-19.5$) appear to be dynamically isotropic and rotationally flattened
(eg Davies et al 1983). Their ``anisotropy'' parameter $(V/\sigma)^*$,
which measures the ratio of the observed value of $(V/\sigma)$ to
the value for an oblate isotropic rotator with the observed
ellipticity, is distributed around a mean of about 1 with small
dispersion.  For the fainter dwarf ellipticals, the mean values of
$(V/\sigma)^*$ rapidly decrease as the magnitudes become fainter.
Many of these fainter systems are found to have low or unmeasurable
rotational velocities along their major axes (Bender \& Nieto 1990;
Bender et al 1991; Geha et al
2002, 2003; Simien \& Prugniel 2002; Pedraz et al 2002; De Rijcke
et al 2001).  Figure 5 of van Zee, Skillman \& Haynes (2004)
shows the dynamical transitions nicely.

The dynamical transition between the intermediate luminosity
ellipticals and the dwarfs is as dramatic and well-defined as the
well-known transition at $M_B\approx -20$ between intermediate
luminosity and giant ellipticals (Davies et al 1983). The
brightest and faintest ellipticals both appear to be flattened by
their anisotropic velocity distributions. These dynamical differences
presumably reflect differences in the formation processes for the
three classes of ellipticals.

While these formation processes are certainly not yet fully understood,
we have at least a working picture for the formation of the two
brighter dynamical classes of ellipticals. It seems likely that
many of the brightest ellipticals form through mergers, which lead
to their anisotropic velocity distribution (eg Naab \& Burkert 
2003), while the intermediate ellipticals are believed to form
dissipatively and so end up rotationally flattened. At this time,
the formation processes for the fainter dwarf ellipticals remain
quite uncertain. They may have several formation routes. Bender
\& Nieto (1990) suggest that the anisotropic velocity distribution
of the fainter dwarf ellipticals is generated by the anisotropic
blowout of their interstellar medium by supernovae during their
evolution and the subsequent dynamical readjustment. The fainter
dwarf ellipticals presumably evolve from some kind of small
star-forming systems, but these progenitor systems must have much
lower angular momentum per unit mass than the rapidly rotating
dwarf irregulars which we observe at the present time.  Our lack of
understanding of the fainter dwarf ellipticals makes it important
to acquire as much information as possible about their dynamics and
content.

The nearby Sculptor (Scl) and Centaurus A (Cen A) groups at mean
distances of about $2.5$ and $3.5$ Mpc contain many dwarf irregular
and dwarf elliptical galaxies. C\^ot\'e et al (1997) made a study
of the gas-rich dwarf irregular galaxies in these two groups,
reaching down to $M_V = -8$, and others have supplemented this
sample (eg Banks et al 1999). At the time, little was known about
the gas-poor dE and dS0 population of these nearby groups. The
fainter dE/dS0 galaxies usually have smooth low surface brightness
light distributions. It is difficult to measure their radial
velocities, and they cannot readily be distinguished from brighter
galaxies at higher redshift on morphological grounds. Images are
needed in which these objects are partly or wholly resolved into
stars.

More recently, Jerjen et al. (1998, 2000a, 2000b) undertook an 
optical program to identify and study a complete sample of the
gas-poor dE and dS0 galaxies in the Scl and Cen A groups. To confirm
that their candidates were dwarf members of these groups, they used
the Surface Brightness Fluctuation (SBF) technique developed by
Tonry \& Schneider (1988) to derive distances for their sample.
Because these groups are so close, Jerjen et al. (1998, 2000a, 2000b)
were able to reach
dwarf galaxies four to five magnitudes fainter than those previously
studied in the Virgo and Fornax clusters, and the groups Leo I,
Dorado, NGC 1400, NGC 5044, and Antlia (eg Ferguson \& Sandage 1990).

Jerjen et al (1998, 2000b) measured SBF distances for 5 faint 
dE/dSph galaxies in each of the Cen A and 
Scl groups. Their absolute magnitudes are in the range M$\rm _B =
-9.5 ~to -15.3$. A few of these objects also have optical velocities
to further confirm their group membership. Table~\ref{dSph-sample}
presents basic data for these 10 early-type dwarf galaxies in the
Scl and Cen A groups.

We are interested in the origin of these dE/dSph galaxies: have they
recently evolved from dIrr galaxies which have expelled their gas
through star formation and stellar winds? It is already known that the
fainter dE and dSph galaxies are not all totally gas-free; some contain
detectable amounts of \hi. For example, among the Local Group dwarf
elliptical galaxies, NGC 147 has not been detected in \hi but NGC 205
and NGC 185 ($M_B = -16.0, -14.7$) have detected \hi masses of about $7
\times 10^5$ \msol~ and $1 \times 10^5$ \msol~ respectively (Young \&
Lo, 1996, 1997).

Some of the fainter dSph galaxies in the Local Group may also contain
detectable \hi. The transition dSph/dIrr galaxy Phoenix (Carignan,
Demers \& C\^ot\'e 1991; St-Germain et al. 1999) has $\sim 10^6$
\msol~ of \hi detected. Observations of the Sculptor dSph
(Carignan et al. 1998; Bouchard, Carignan \& Mashchenko 2003), at
a distance of 80 kpc, showed $2 \times 10^5$ \msol~ of \hi
located in two clouds displaced from its optical center, suggesting
expulsion from the galaxy or possibly residual gas from an earlier
star-forming stage. It is not yet clear whether this residual gas
is in any way associated with the extended star formation history
of this dSph galaxy (eg Grebel 1998). Blitz \& Robishaw (2000)
suggested that many of the Local Group dSph and dSph/dIrr galaxies
contain significant masses of \hi, often displaced by a few kpc
from the centers of the galaxies. Many of the dSph galaxies are
projected on fairly dense fields of small high velocity clouds
(Putman et al 2002), and it is difficult to tell whether the \hi
near these galaxies is just a foreground HVC or is really part of
the galaxy. A recent study of \hi in the Local Group dSph galaxies
by Bouchard, Carignan and \& Staveley-Smith (2005) finds that only 
two (Scl and Tucana) can be distinguished from high velocity
clouds at a fairly high level of confidence.

Surveys of galaxy clusters (Virgo: Binggeli et al. 1985;
Fornax: Ferguson 1989; Centaurus: Jerjen \& Dressler 1997) have 
revealed large
populations of dwarf ellipticals relative to the numbers in groups
and the field. The cluster environment is clearly conducive to the
formation of dwarf ellipticals, presumably by the transformation
of other kinds of galaxies via harassment, ram pressure stripping
and blowout. Conselice et al (2003) have compiled a sample of \hi
observations of Virgo cluster dE galaxies. Of the 48 dE galaxies
observed, 5 were convincingly detected, with \hi masses $\sim$ a
few $\times 10^7$ \msol. These 5 are all near the periphery of
the cluster. The authors argue that they are probably gas-rich
systems recently accreted by the cluster, and are in the process
of transformation by stripping.

\begin{figure}
\epsscale{0.6}
%\plotone{8PANELS.ps}
\plotone{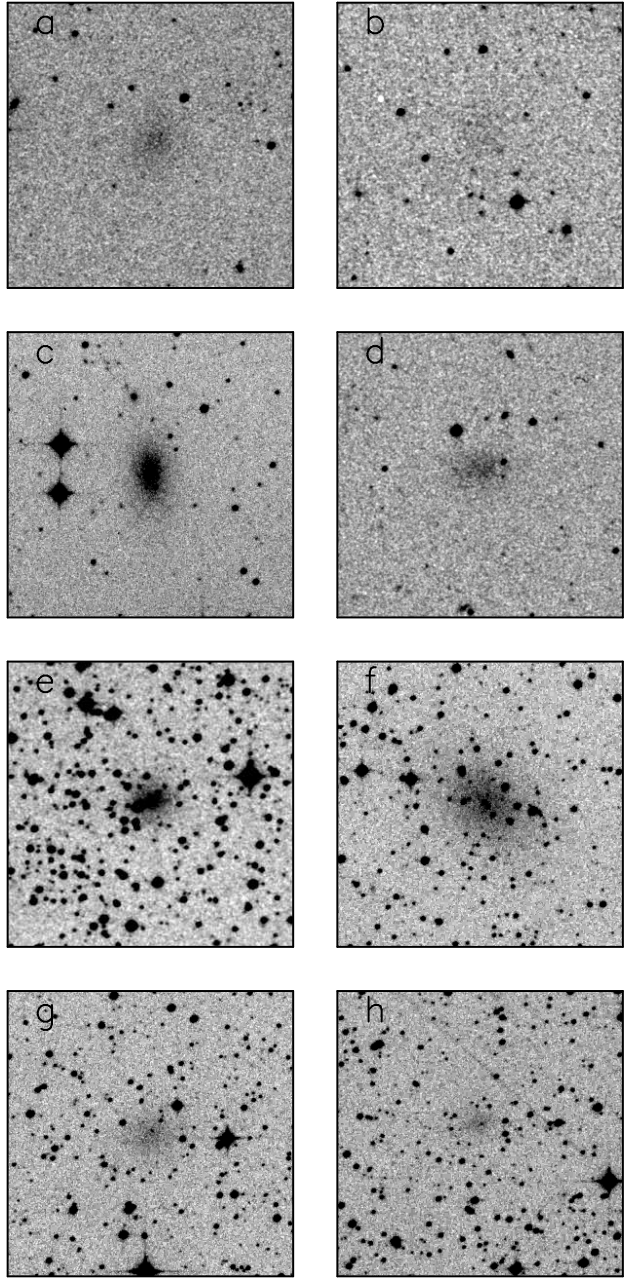}
\caption{STScI DSS optical images of the 8 non-detections. The galaxies are: (a)
ESO540-032, (b) SC22, (c) ESO294-010, (d) ESO540-030, (e) ESO219-010, 
(f) ESO269-066, (g) AM1339-445, and (h) AM1343-452. East is
left and North is up. The FOV is $5\arcmin \times 5\arcmin$.
\label{DSS1}}
\end{figure}

\begin{figure}
%\plotone{2PANELS.ps}
\plotone{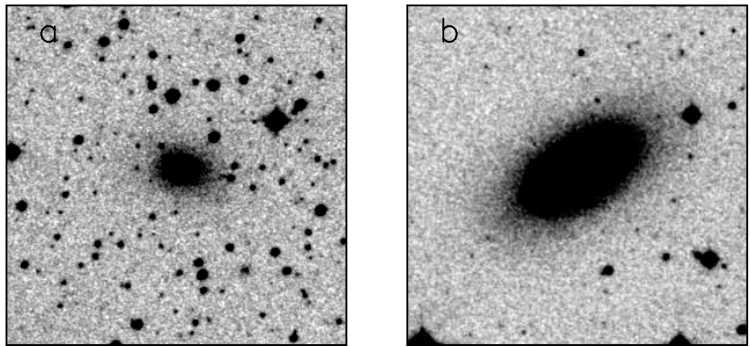}
\caption{STScI DSS optical images of the 2 detections. The galaxies are: (a)
ESO384-016, and (b) NGC 59. East is left and
North is up. The FOV is $5\arcmin \times 5\arcmin$.
\label{DSS2}}
\end{figure}

The Local Group dSph galaxies studied by Bouchard et al (2005) are
mostly fainter than $M_B = -12$. The Virgo dE sample compiled by
Conselice et al (2003) are mostly brighter than $M_B = -14$. In the
absolute magnitude gap between the Virgo and Local Group samples,
the Jerjen et al surveys of the nearby Cen~A and Scl groups provide
a sample of 10 dE galaxies with $M_B = -9$ to $-15$ which we have
observed at 21cm with the ATCA. Only two were detected in \hi and
observed with the Green Bank Telescope (GBT), along with one non-detection.
These two detections (ESO384-016 in Cen A and NGC 59 in Scl) are
the subject of this paper. Images of the eight non-detected galaxies
and the two detected galaxies are shown in Figure~\ref{DSS1} and 
Figure~\ref{DSS2} respectively.

\section{Observations and Reduction}

The \hi observations for our ten galaxies were made with the Australia
Telescope Compact Array (ATCA) in the 750 m configuration. The total
integration time per galaxy was $\sim 12$ hours, and included 10 min on
a primary calibrator, and then a sequence of 5 min on a secondary
calibrator and 50 min on source. For the two detected galaxies, the
observations were taken  on 1998 April 11 for ESO384-016, and
on 1998 July 28 for NGC 59. Details of the observations can be found in
Table~\ref{dSph-obs}. The noise levels for the observations of the
undetected galaxies are similar to those given in the last row of
Table~\ref{dSph-obs}. 

Data reduction was conducted using the Multichannel Image Reconstruction,
Image Analysis and Display package (MIRIAD/ATNF) (Sault et al. 1995) for 
calibrating the data. The NRAO package AIPS was used to extract the moments
maps i.e. the \hi column density, velocity field and velocity
dispersion.

Both our galaxies suffer from an elongated synthesised beam because
the ATCA is an E-W array (for these observations). The N-S resolution
of our beam is degraded by a factor of $\rm 1/sin(dec)$ relative
to the E-W resolution. For our two galaxies, this corresponds to
beam profiles of $167\arcsec \times 61\arcsec$ for NGC 59, and
$91\arcsec \times 58\arcsec$ for ESO384-016, for our maps
produced with natural weighting. Although it is customary
to convolve the data with a circular beam, in our case, this strategy
would mean sacrificing a significant amount of resolution in the
E-W direction. This could potentially lead to unnecessary loss of
information for small sources like our dwarf galaxies.

The three most luminous dwarfs in our sample, including ESO384-016
and NGC 59, were subsequently observed in the 21cm line with the
100m Green Bank Telescope (GBT) of the NRAO at an angular resolution
of 9 arcmins. The receiver system temperature
at zenith was 18 K, but because these galaxies are quite far south
(ESO 269-066 was observed at an elevation of $6\deg$),
emission from the atmosphere added between 5 and 20 K to the system
temperatures. The spectra covered about 1000 km s$^{-1}$ at a
velocity resolution of 1.2 km s$^{-1}$. Each galaxy was observed
for 5 to 10 minutes in frequency-switched mode;
a low-order polynomial was fit to each spectrum to
remove residual instrumental baseline.

\section{\hi Content and Distribution}

\begin{figure}
%\plotone{NGC59HI.ps}
\plotone{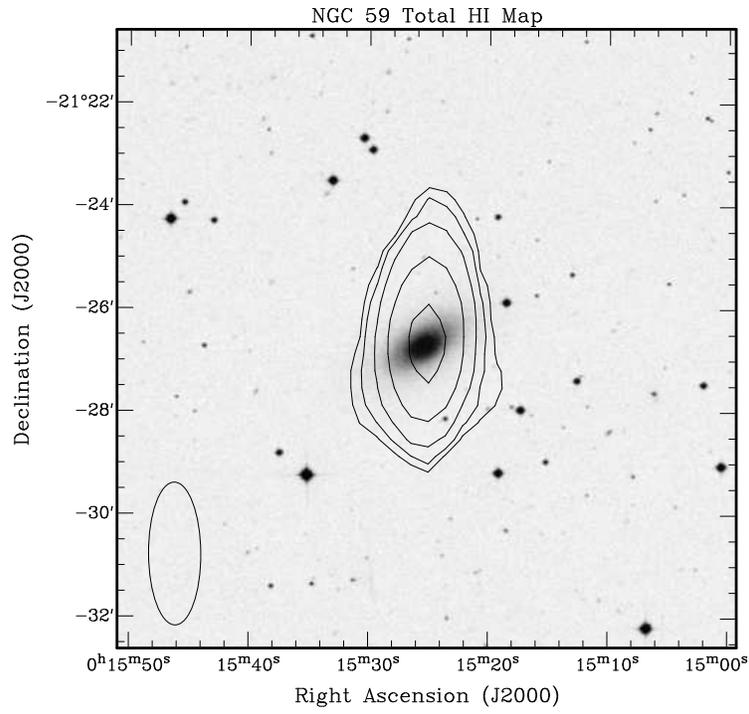}
\caption{Total \hi map of NGC 59 superposed on the STScI DSS
optical image using the displaying package KARMA (Gooch 1997). 
The contours are $\rm 0.8, 1.6, 3.2, 6.4
~ and~ 12.8 \times 10^{19}~ cm^{-2}$. 
\label{HINGC}}
\end{figure}

\begin{figure}
%\plotone{ESO384016HI.ps}
\plotone{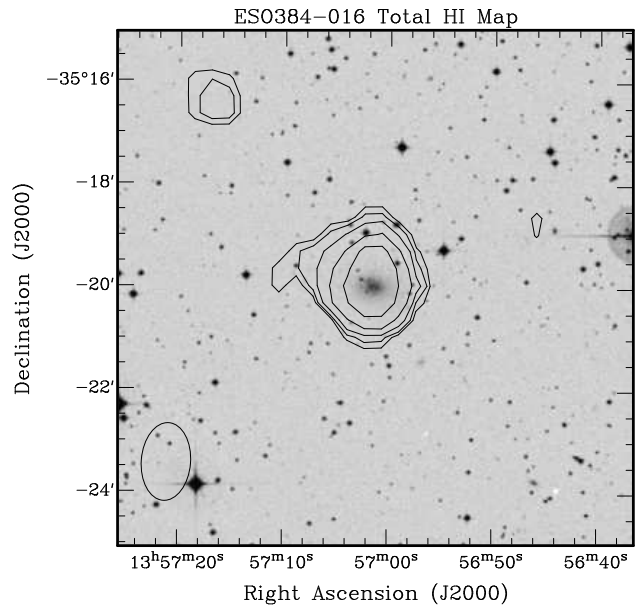}
\caption{Total \hi map of ESO384-016 superposed on the STScI DSS
optical image using the displaying package KARMA (Gooch 1997). 
The contours are $\rm 0.5, 1.0, 2.0, 4.0,~ and~ 8.0 
\times 10^{19}~ cm^{-2}$.
\label{HIESO}}
\end{figure}

\begin{figure}
%\plotone{newGBT3GAL.ps}
\plotone{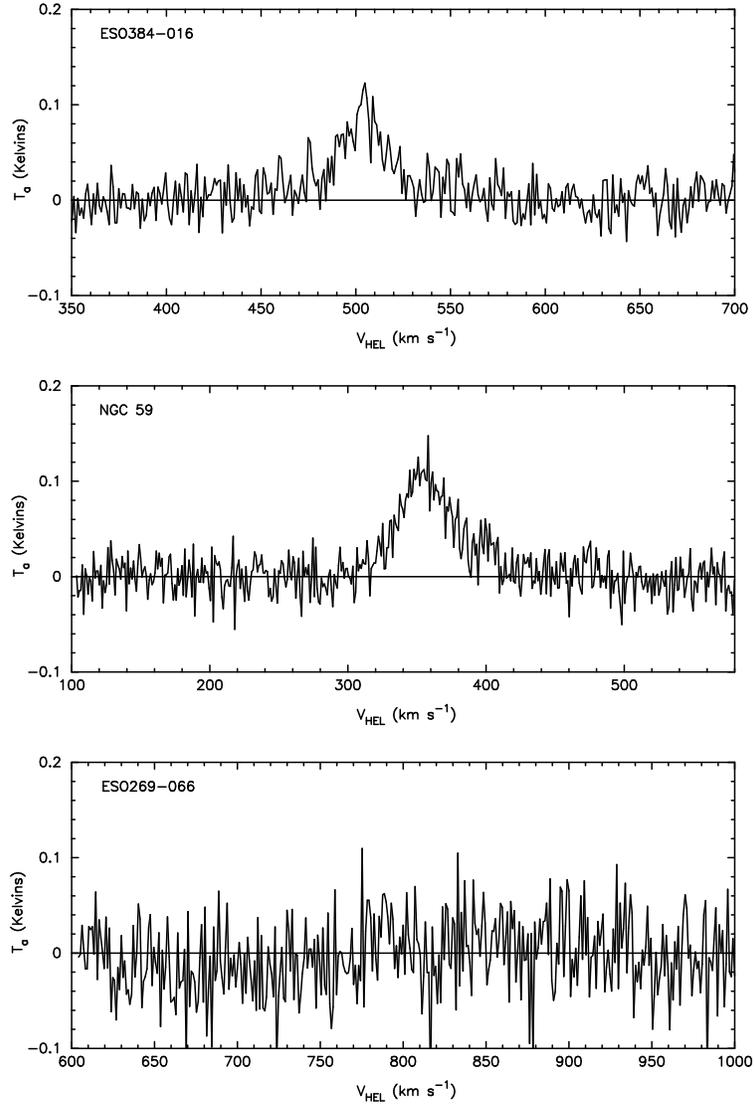}
\caption{Single dish spectra from the GBT.
\label{GBT}}
\end{figure}

The \hi distributions of NGC 59 and ESO384-016 are shown in
Figure~\ref{HINGC} and Figure~\ref{HIESO}. From the comparison with
the beam shown in the bottom left corner of the maps, it can be seen
that both are barely resolved. 
Once detected with the ATCA, NGC 59 and ESO0384-016 were
also observed with the GBT, along with ESO269-066 which was an ATCA
non-detection. The GBT observations gave similar results: the GBT
spectra are shown in Figure~\ref{GBT}.

Table~\ref{dSph-results} summarizes the velocity 
(heliocentric) data and gives the \hi
masses.  For NGC 59, the optical and \hi observations give very similar
systemic velocities $\sim$358 \kms. For ESO384-016, while the 2 \hi
velocities give a similar result of 504 \kms, the optical velocity is
nearly 60 \kms higher. Optical observations of such low surface
brightness systems are difficult, and the \hi determinations are surely
more accurate: the \hi velocities are adopted.

From the GBT  single dish profiles, total fluxes of 1.53 Jy \kms and
2.72 Jy \kms and line widths (FWHM) of 33.7 $\pm 2.3$ \kms and 50.6
$\pm 2.1$ \kms were measured for ESO384-0166 and NGC 59, respectively.
A mean of the ATCA and the GBT fluxes gives \hi masses of $6.0 \pm 0.5
\times 10^6$ \msol for ESO384-016 and $1.4 \pm 0.1 \times 10^7$ \msol 
for NGC 59. From the noise in the ATCA maps and assuming a line width of
40 \kms ( $\approx$ mean for our two detections), we can put an upper
(3-$\sigma$) limit of $\sim 5.3 \times 10^5$ \msol on the \hi content
of the non-detected objects in Sculptor ($\Delta \sim 2.5$ Mpc) and
$\sim 1.0 \times 10^6$ \msol in the Centaurus group ($\Delta \sim 3.5$
Mpc).

\section{Discussion and Conclusion}

As we have seen, the two galaxies that were detected are dS0s while
all the other galaxies in the sample are dEs. Looking at their optical
properties in Jerjen et al. (2000a), it can be seen that those two
types are quite different, e.g. their surface brightness profiles (see
Fig. 3 of Jerjen et al. 2000a). While the profiles for dEs tend to
flatten in the very inner parts, those of dS0s stay linear all the way
to the center and sometimes even exhibit what looks like a bulge
component. Their best-fitting S\'ersic profiles give S\'ersic n values
for our 2 dS0s of n $\sim$ 0.7 (n = 0.25 for a de Vaucouleurs 
profile) , while for the dEs $1.2 < n < 1.7$.
This change of slope in the gradient of the radial light distribution
is what was used by Sandage and Binggeli (1984) to define the dS0
class.  Later, Binggeli \& Cameron (1991) gave five main criteria to
classify a system as dS0 instead of dE: ({\it i}) unusual King
parameters; ({\it ii}) high flattening; ({\it iii}) presence of a lens;
({\it iv}) asymmetry (change of ellipticity, presence of a bar,
isophotal twists, boxiness); and ({\it v}) central irregularity
(clumpiness).

The present study suggests another difference: the high \hi content.
While for the few nearby dEs where \hi is detected, such as NGC 185 and
NGC 205 (Young \& Lo 1997), the \mhil is very small (0.004 and 0.003,
respectively), the \mhil of ESO384-016 and NGC 59 are much larger at
0.21 and 0.07, respectively.  Those \mhil are similar to what is seen
in dSph/dIrr galaxies such as Phoenix (\mhil $\sim 0.21$; St-Germain et
al. 1999).  In terms of their \hi content, they are closer to dIrrs
such as GR 8 (Carignan et al. 1990) and Sextans A (Skillman et al.
1988) which have \mhil $\sim 1.0$ than to dEs. In fact, their \mhil is
similar to the \hi content seen in a normal spiral like the MW (\mhil
$\sim 0.18$).

Looking at the total \hi maps, not much can be said about NGC 59 since
it is not really resolved. However, the resolution is slightly better
for ESO384-016. Fig. 4 indicates that the contours are somewhat
compressed on the west side with an elongated  component extending to
the east. This is what we might expect if this system is falling toward
the center of the Centaurus Group and was in interaction with its
intergalactic medium (IGM). ESO384-016 lies on the east side of the
group, so the orientation of these features is in the expected
direction. This is reminescent of what is seen in Holmberg II (Bureau
\& Carignan 2002) which is thought to be falling toward the center of
the M81 Group.

\begin{figure}
%\plotone{DSPHMB.ps}
\plotone{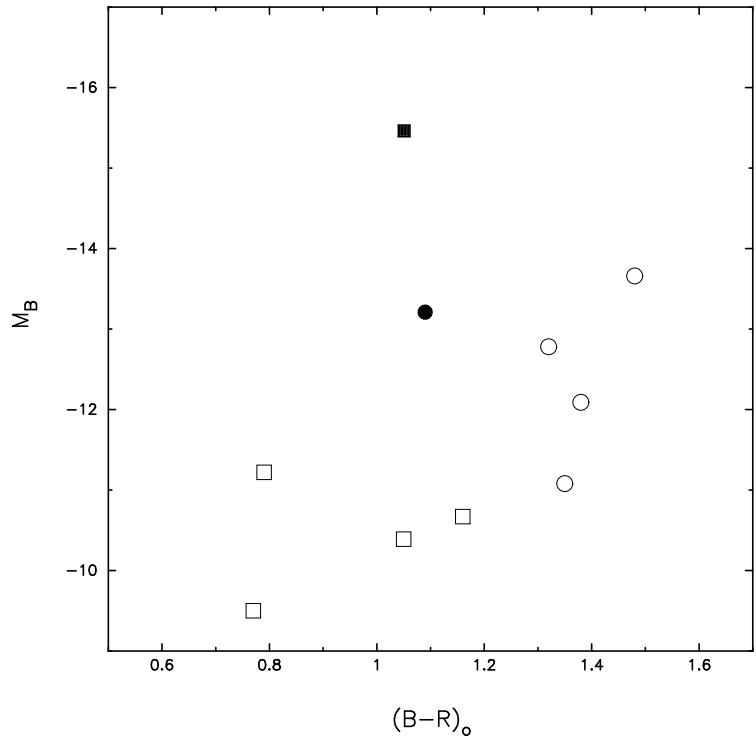}
\caption{Color--Magnitude diagram for our sample. Squares are for Sculptor
objects, circles for Centaurus and filled symbols are the detected dS0 galaxies.
\label{CMD}}
\end{figure}

The color-magnitude diagram for the galaxies in our sample is shown in
Figure~\ref{CMD}. The two detected dS0 are amongst the brightest, with
$M_B < -13$. The galaxies in the two groups are well separated in
color, with the Sculptor systems having $(B - R)_{T}^{0} \leq 1.1$ and the
Centaurus galaxies $(B-R)_{T}^{0} \geq 1.1$.  It is more interesting to
look at the radial color profiles of the galaxies in the sample (Fig. 4
of Jerjen et al. 2000a). The two detected galaxies are those with the
largest color gradient, with $(B - R)_{T}^{0}$ varying from 0.60-0.85 in the
center to $\sim 1.5$ in the outer parts. While the colors in the outer
parts are typical of E-S0 galaxies, the colors in the center are more
like what is seen in late-type spiral and irregular galaxies. This may
indicate that recent star formation has taken place in the center of
those two galaxies using some of the gas from the detected \hi
reservoirs.

\begin{figure}
%\plotone{VIRGOHIST.ps}
\plotone{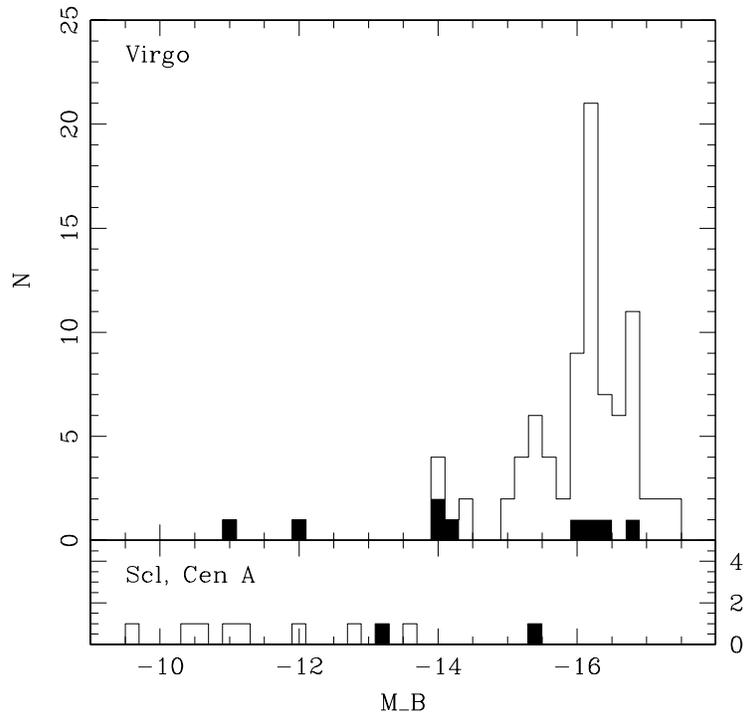}
\caption{Distribution of the \hi detected dwarfs in the Virgo sample of 
Conselice et al (2003) and in our Scl/Cen A sample, as a function
of absolute magnitude. 
\label{HISTO}}
\end{figure}

The only other sample with which our results could be compared to is 
the Conselice et al. (2003) Virgo dE sample. The distribution of the two
samples as a function of absolute magnitude is shown in
Figure~\ref{HISTO}. It can be seen that while our objects are mostly
fainter than $M_B \sim -14$ (9/10), most of the Virgo galaxies are
around $M_B \sim -16$. However, our \mhil are only slightly smaller
than their mean \mhil of $0.36 \pm 0.16$ while
their mean \hi mass is a factor of $\sim 10$
larger at $<$\mhi$>$ $= 2.8 \times 10^8$ \msol (Figures 8 \& 9). 
Comparing the location
of our two detected galaxies (Figures 10 \& 11) to the
detected systems in Virgo (Fig. 5 of Conselice et al. 2003) we see
that, as in Virgo, our \hi-detected systems are found near the
periphery of the groups.

The main results of this paper are:
\begin{itemize}
\item Of the 10 dE/dSO of the Sculptor and Centaurus groups observed with the 
ATCA, only the two dS0s ESO384-016 and NGC 59 were detected at 21cm with 
our detection limit of $\sim 1.0 \times 10^6$ \msol for the Centaurus group 
and $\sim 5.3 \times 10^5$ \msol for the Sculptor group. These two detections
were subsequently confirmed with GBT observations.
\item The two detected systems also happen to be among the most luminous
early-type dwarf galaxies in the two groups with $M_B =$ --15.3 \& --13.2
\item We find \hi masses of $1.4 \pm 0.1 \times 10^7$ \msol~ for NGC 59 and
$6.0 \pm 0.5 \times 10^6$ \msol~ for ESO384-016.
\item The \hi properties (mass, \mhil) of NGC 59 and ESO384-016 
are much closer 
to those of dIrrs or normal spirals than of dEs.
\item Features in the \hi distribution  of ESO384-016 suggest that it may be 
experiencing
ram pressure stripping from the Centaurus Group IGM while falling toward the 
centre of the group.
\item The $(B-R)_{T}^{0}$ colors of the two detected galaxies suggest
that recent star formation has taken place from the reservoir of detected gas.
\item Our two detected galaxies appear to have similar \mhil 
than the ones of the detected dwarfs in the Virgo 
cluster and, as in Virgo, they are located on the outskirt of the groups.
\end{itemize}

As in the Local Group, we expect that more sensitive observations would
succeed in detecting HI reservoirs in or close to dEs (dS0s, dSph,
dSph/dIrr) in nearby groups. We know from population studies that, in
some of those systems, \hi is needed to explain the presence of the
observed young or intermediate age populations
(Grebel et al 2003). The two systems
presented here were detected after only 5-10 minutes observations with
the GBT.  With the much improved radio receivers (T$_{sys} \leq 20$ K),
it is surely worthwhile to do much deeper observations of those
galaxies.  The knowledge of their ISM content and kinematics are
essential ingredients if we want to understand the evolution of dwarf
systems and pin down the progenitors of the present-day dEs.

\acknowledgments

SFB wishes to acknowledge financial support from the Institute of
Astronomy, Cambridge (through grants from PPARC), for financing all the 
ATCA observing runs of this project. SFB is also grateful for the assistance
from Tasso Tzioumis for observing one galaxy during a period of flooded
roads, Kate Brooks, Haida Liang, St\'ephanie C\^ot\'e and Bob Sault for
their help and discussion. CC acknowledges support from NSERC (Canada).
Based on photographic data obtained using the UK Schmidt Telescope.
The UK Schmidt Telescope was operated by the Royal Observatory
Edinburgh, with funding from the UK Science and Engineering Research
Council, until 1988 June, and thereafter by the Anglo-Australian
Observatory. Original plate material is copyright (c) the Royal
Observatory Edinburgh and the Anglo-Australian Observatory. The
plates were processed into the present compressed digital form with
their permission. The Digitized Sky Survey was produced at the Space
Telescope Science Institute under US Government grant NAG W-2166.
The National Radio Astronomy Observatory is operated by Associated 
Universities, Inc., under a cooperative agreement with the
National Science Foundation.

\clearpage
\newpage

%%%%%%%%%%%
%References
%

\clearpage
\newpage

%%%%%%%%%%%%%%%%%%%%%
% Inclusion of Tables
%
%Table 1

\begin{deluxetable}{lcccccccc}
\tabletypesize{\scriptsize}
%\tablecolumns{9}
%\tablewidth{0pc}
\tablewidth{0pt}
\tablecaption{Parameters of the galaxies in the sample (Jerjen et al. 1998, 2000b) 
\label{dSph-sample}}
\tablehead{
\colhead{} &
\colhead{} &
\colhead{R.A.} &
\colhead{Decl.} &
\colhead{$M_{B}^{0}$} &
\colhead{$(B-R)_{T}^{0}$} &
\colhead{D} &
\colhead{$V_{opt}$\tablenotemark{a}} &
\colhead{$V_{HI}$\tablenotemark{ab}} \\
\colhead{Galaxy} &
\colhead{Type} &
\colhead{(J2000)} &
\colhead{(J2000)} &
\colhead{(mag)} &
\colhead{(mag)} &
\colhead{(Mpc)} &
\colhead{(\kms)} &
\colhead{(\kms)} \\
\colhead{(1)} &
\colhead{(2)} &
\colhead{(3)} &
\colhead{(4)} &
\colhead{(5)} &
\colhead{(6)} &
\colhead{(7)} &
\colhead{(8)} &
\colhead{(9)}}
\startdata
Centaurus A Group & & & & & & & & \\
ESO219-010 & dE,N & 12~ 56~ 09.6~ & $-50~ 08~ 38$ & -12.78 & 1.32 & 4.8(0.4)
& \nodata & \nodata \\
ESO269-066 & dE & 13~ 13~ 09.5~ & $-44~ 52~ 56$ & -13.66 & 1.48 & 4.0(0.5)
& 784(32) & \nodata \\
AM1339-445 & dE & 13~ 42~ 05.8~ & $-45~ 12~ 21$ & -12.09 & 1.38 & 3.7(0.5)
& \nodata & \nodata \\
AM1343-452 & dE & 13~ 46~ 17.8~ & $-45~ 41~ 05$ & -11.08 & 1.35 & 4.0(0.7) 
& \nodata & \nodata \\
ESO384-016 & dS0 & 13~ 57~ 01.2~ & $-35~ 19~ 59$ & -13.21 & 1.09 &4.2(0.1)
& 561(32) & \nodata \\
Sculptor Group & & & & & & & & \\
NGC59 & dS0 & 00~ 15~ 25.1~ & $-21~ 26~ 38$ & $-15.30$ & 1.07 & 4.4(0.2) 
& 362(10) & 367 (357 in \ha)\\
SC22\tablenotemark{c} & dE & 00~ 23~ 51.7~ & $-24~ 42~ 18$ & $-9.50$ & 0.79
& 2.7(0.2) & \nodata & \nodata \\
ESO294-010 & dS0/Im & 00~ 26~ 33.4~ & $-41~ 51~ 19$ & $-10.67$ & 1.17 &1.7
(0.1) & 117(5) & \nodata \\
ESO540-030 & dE/Im & 00~ 49~ 21.1~ & $-18~ 04~ 34$ & $-11.22$ & 0.83 & 3.2
(0.1) & \nodata & \nodata \\
ESO540-032& dE/Im & 00~ 50~ 24.5~ & $-19~ 54~ 23$ & $-10.39$ & 1.08 & 2.2
(0.1) & \nodata & \nodata \\
\enddata
\tablenotetext{a}{Heliocentric velocities}
\tablenotetext{b}{Parkes HI single dish velocity from C\^ot\'e et al. 1997}
\tablenotetext{c}{a.k.a. Scl-dE1 (Jerjen et al. 1998) }
\end{deluxetable}

\clearpage
\newpage

%Table 2

\begin{deluxetable}{lcc}
\tablecolumns{3}
\tablewidth{0pc}
\tablecaption{\hi Observations for ESO384-016 and NGC 59 \label{dSph-obs}}
\tablehead{
\colhead{} &
\multicolumn{1}{c}{ESO384-016} &
\multicolumn{1}{c}{NGC 59}} 
\startdata
Array Configuration & 750A & 750E \\
Receiver (cm) & 20 & 20  \\
Integration Time (hr) & 11.68 & 11.47  \\
Transition (MHz) & \hi 1420 & \hi 1420  \\
Channel Bandwidth (kHz) & 15.6 & 15.6  \\
Velocity Increment (\kms) & 3.3 & 3.3  \\
Number channel per baseline & 512 & 512 \\
Correlator Configuration & FULL & FULL  \\
Primary Calibrator& PKS1934-638 & PKS1934-638  \\
Secondary Calibrator & PKS1320-446 & PKS0023-263  \\
Bandwidth (MHz) & 8 & 8  \\
Cell size (arcsec) & 20 & 20 \\
RMS noise (mJy/beam) & 3.0 & 2.6 \\
$3\sigma$/channel ($\rm cm^{-2}$) & $7 \times 10^{18}$ & $3 \times 10^{18}$ \\
\enddata
\end{deluxetable}

\clearpage
\newpage

%Table 3 

\begin{deluxetable}{lccccccccc}
\tablecolumns{10}
\tablewidth{0pc}
\tablecaption{\hi results for ESO384-016 and NGC 59 \label{dSph-results}}
\tablehead{
\colhead{} &
\multicolumn{3}{c}{Systemic V$_{hel}$ (\kms)} & 
\colhead{} &
\multicolumn{3}{c}{$M_{HI}$~($M_\odot$)} & 
\colhead{} &
\multicolumn{1}{c}{$M_{HI} / L_B$} \\
\cline{2-4} \cline{6-8} \cline{10-10} \\ 
\colhead{} &
\colhead{Optical} &
\colhead{ATCA} &
\colhead{GBT} &
\colhead{} &
\colhead{ATCA} &
\colhead{} &
\colhead{GBT} &
\colhead{} &
\colhead{}} 
\startdata
ESO384-016 & $561$ & $504$ & $504$ &  & $5.6 \pm 0.3 \times 10^6$ & & $6.5 \pm 0.1 \times 10^6$ & & 0.21 \\
NGC 59 & $362$ & $359$ & $357$ & & $1.5 \pm 0.3 \times 10^7$ & & $1.3 \pm 0.1 \times 10^7$ & & 0.07 \\
\enddata
\end{deluxetable}

\clearpage
\newpage

%%%%%%%%%%%%%%%%%%%%%%
% Inclusion of Figures
%
% Figure containing the DSS images of the 8 non-detections
%

\clearpage
\newpage

%
% Figure containing the DSS images of the 2 detections
%

\clearpage
\newpage

%
% Figure containing the total HI map of NGC 59.
%

\clearpage
\newpage

%
% Figure containing the total HI map of ESO384-016.
%

\clearpage
\newpage

%
% Figure containing the GBT spectra.
%

\clearpage
\newpage

%
% Figure containing the CMD of the sample
%

\clearpage
\newpage

%
% Figure containing the histograms
%

\clearpage
\newpage

\begin{figure}
\plotone{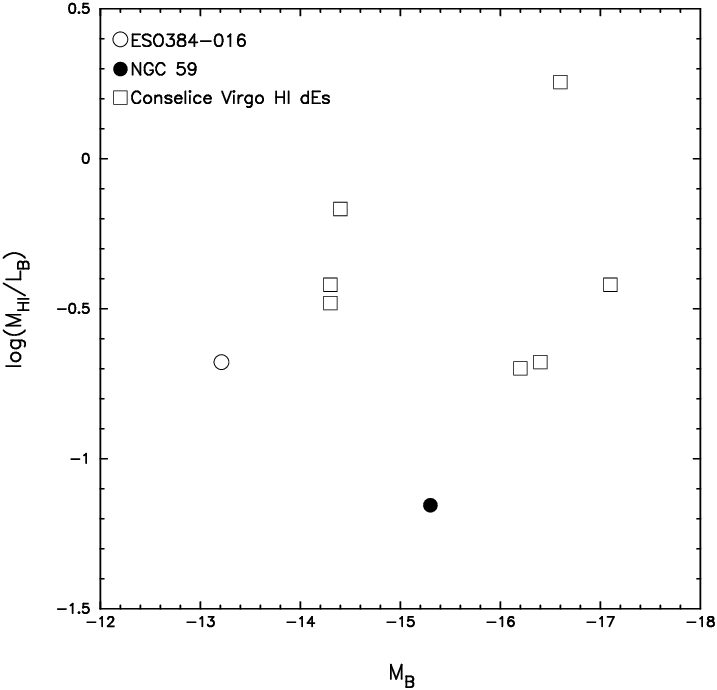}
\caption{Absolute magnitude ($M_B$) vs. \mhil~ diagram for Virgo \hi
detected dE galaxies (Conselice et al 2003) and the \hi detected dwarf 
galaxies in this study.}
\end{figure}

\clearpage
\newpage

\begin{figure}
\plotone{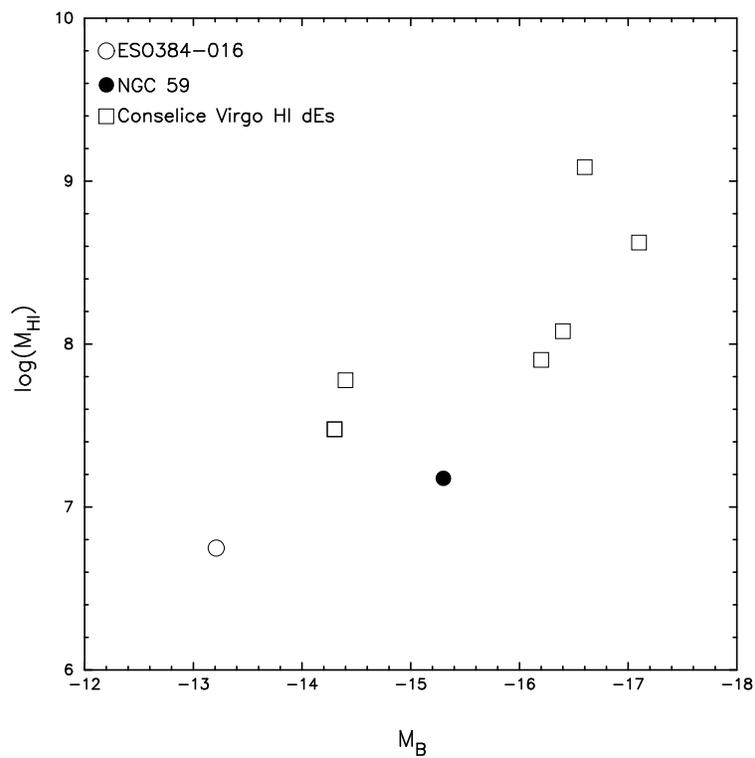}
\caption{Absolute magnitude ($M_B$) vs. \mhi~ diagram for Virgo \hi
detected dE galaxies (Conselice et al 2003) and the \hi detected dwarf 
galaxies in this study.}
\end{figure}

\clearpage
\newpage

\begin{figure}
\plotone{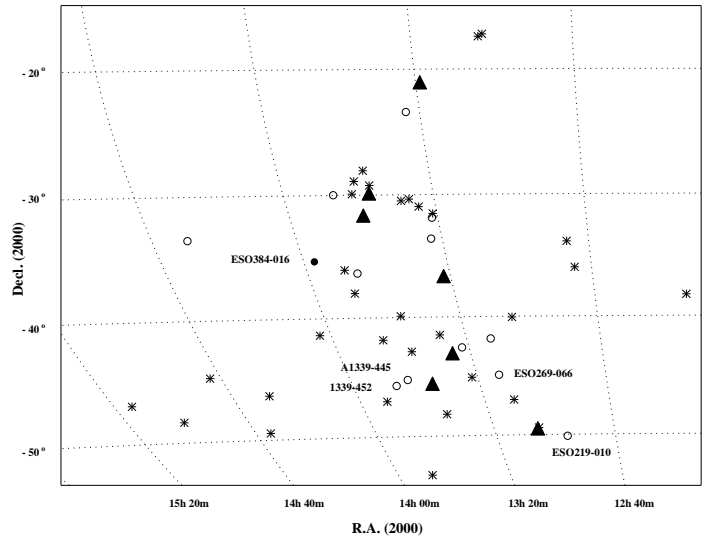}
\caption{Distribution in R.A. (J2000.0) and Decl. (J2000.0) of the 
Cen A group. Filled triangles: major group members, stars: late-type 
dwarf galaxies (C\^ot\'e et al 1997, Jerjen et al 2000b), unlabelled
circles: early-type galaxies (Jerjen et al 2000a), labelled
circles: early-type galaxies observed in this study, labelled
filled circle: early-type galaxy ESO384-016 observed in
this study with HI detected.}
\end{figure}

\clearpage
\newpage

\begin{figure}
\plotone{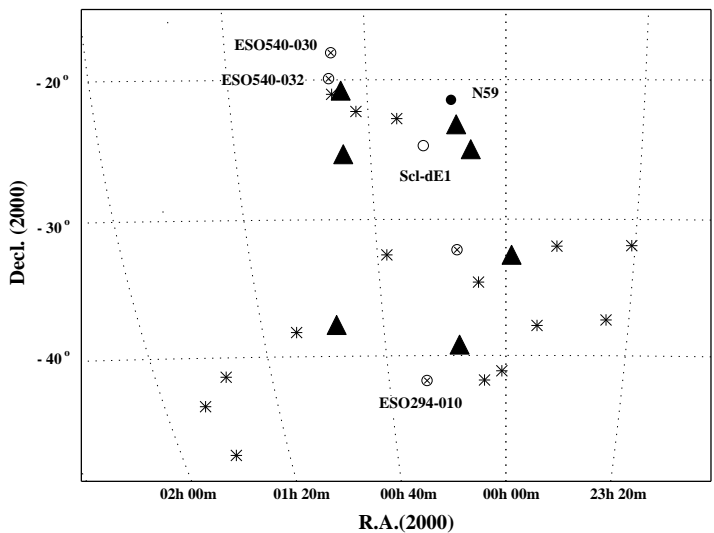}
\caption{Distribution in R.A. (J2000.0) and Decl. (J2000.0) of the
Scl group. Filled triangles: major group members, stars: late-type
dwarf galaxies (C\^ot\'e et al 1997), circle: early-type
galaxy Scl-dE1 (SC22) observed in this study, but not
detected in HI, labelled filled circle: early-type galaxy
NGC59 observed in this study with HI detected, labelled
crossed circle: early-type galaxies observed in this study
with no HI detected at our detection limit of \mhi = 
$\approx 10^7$ \msol, but HI detected at the \mhi = $\sim 10^{(5-6)}$
\msol level (Bouchard, Jerjen, Da Costa \& Ott, 2005), unlabelled
crossed circle: early-type galaxies ESO410-005, not observed
in our study but detected in HI by Bouchard, Jerjen, Da Costa \& Ott (2005).}
\end{figure}

\end{document}